\documentclass[vecphys]{svmult}

\usepackage{makeidx}
\usepackage{graphicx}
\usepackage{multicol}
\usepackage[bottom]{footmisc}
\usepackage{amsmath}
\usepackage{amsfonts}
\usepackage{graphics}
\usepackage{subfigure}
\usepackage{epsfig}
\usepackage{rotate}
\usepackage{ifpdf}
\usepackage{graphicx,amssymb}

\makeindex
\begin{document}

\title*{Dynamical study of 2D and 3D barred galaxy models}

\author{T.~Manos\inst{1,2} \and E.~Athanassoula\inst{2}}

\institute{ Center for Research and Applications of
Nonlinear Systems (CRANS), Department of Mathematics, University of
Patras, GR--26500, Greece.\\
\texttt{thanosm@master.math.upatras.gr} \and Laboratoire
d'Astrophysique de Marseille (LAM), Observatoire Astronomique \\de
Marseille-Provence (OAMP), 2 Place Le Verrier, 13248 Marseille,
C\'edex \\04, France.\\ \texttt{lia@oamp.fr}}

\maketitle
\begin{abstract}
\label{abst}

We study the dynamics of 2D and 3D barred galaxy analytical models,
focusing on the distinction between regular and chaotic orbits with
the help of the Smaller ALigment Index (SALI), a very powerful tool
for this kind of problems. We present briefly the method and we
calculate the fraction of chaotic and regular orbits in several
cases. In the 2D model, taking initial conditions on a Poincar\'{e}
$(y,p_y)$ surface of section, we determine the fraction of regular
and chaotic orbits. In the 3D model, choosing initial conditions on
a cartesian grid in a region of the $(x, z, p_y)$ space, which in
coordinate space covers the inner disc, we find how the fraction of
regular orbits changes as a function of the Jacobi constant.
Finally, we outline that regions near the $(x,y)$ plane are
populated mainly by regular orbits. The same is true for regions
that lie either near to the galactic center, or at larger relatively
distances from it.
\end{abstract}

\vspace{-0.5 cm}
\section{Introduction}
\label{Intro}

The dynamical evolution of galactic systems depends crucially on their
orbital structure and in particular on what fraction of their orbits
is regular, or chaotic. Thus to permit further studies, it is essential to
be able to distinguish between these two types of orbits in a manner
that is both safe and efficient. This is not trivial and becomes yet
more complicated in systems of many degrees of freedom. A
summary of the methods that have been developed over the years
can be found in \cite{Con_spr}.

In the present paper we use a method based on the properties of two
deviation vectors of an orbit, the ``Smaller ALingment Index" (SALI)
\cite{sk:1}. It has been applied successfully in different dynamical
systems
\cite{sk:1,sk:3,sk:5,Pan:1,Ant:2,Ant:3,Bou:1,Man:1,Man:2,Man:3,Man:4},
frequently also under the name Alignment Index (AI)
\cite{VKS1,VKS2,VHC,KVC,KEV} and has been shown to be a fast and
easy to compute indicator of the chaotic or ordered nature of
orbits. We first recall its definition and we then show its
effectiveness in distinguishing between ordered and chaotic motion
by applying it to a barred potential of 2 and 3 degrees of freedom.
Recently, a generalization of the SALI, the ``Generalized ALignment
Index" (GALI) was introduced by Skokos et al. (2007) \cite{sk:6},
which includes the full set of the  $k$ initially linearly
independent deviation vectors of the system to determine if an orbit
is chaotic or not.

\vspace{-0.5 cm}
\section{Definition of the Smaller ALigment Index - SALI}

Let us consider the  $n$--dimensional phase space of a conservative
dynamical system, which could be a symplectic map or a Hamiltonian
flow. We consider also an orbit in that space with initial condition
$P(0)=(x_{1}(0),x_{2}(0),...,x_{n}(0))$ and two deviation vectors
$\vec{w}_{1}(0)$, $\vec{w}_{2}(0)$ from the initial point $P(0)$. In
order to compute the SALI for a given orbit one has to follow the
time evolution of the orbit itself, as well as
two deviation vectors $\vec{w}_{1}(t),\vec{w}_{2}(t)$ which
initially point in two different directions. At every time step the
two deviation vectors $\vec{w}_{1}(t)$ and $\vec{w}_{2}(t)$ are
normalized by setting:

\begin{equation}\label{norm_dev}
\hat{w}_{i}(t)=\frac{\vec{w}_{i}(t)}{\|\vec{w}_{i}(t)\|}, \quad
i=1,2
\end{equation}

\noindent
and the SALI is then computed as:

\begin{equation}\label{eq:SALI}
    SALI(t)=min \left\{\left\|\hat{w}_{1}(t)+\hat{w}_{2}(t)\right\|,\left\|\hat{w}_{1}(t)-\hat{w}_{2}(t)\right\|\right\}.
\end{equation}

The properties of the time evolution of the SALI clearly distinguish
between regular and chaotic motion as follows: In the case of
Hamiltonian flows or  dimensional symplectic maps with $n \geq 2$,
the SALI fluctuates around a non--zero value for regular orbits
\cite{sk:1,sk:3}. In general, two different initial deviation
vectors become tangent to different directions on the torus,
producing different sequences of vectors, so that SALI does not tend
to zero but fluctuates around positive values. On the other hand,
for chaotic orbits SALI tends exponentially to zero. Any two
initially different deviation vectors tend to coincide in the
direction defined by the nearby unstable manifold and hence either
coincide with each other, or become opposite.

\begin{figure}[t]
\centering
\includegraphics[height=6cm]{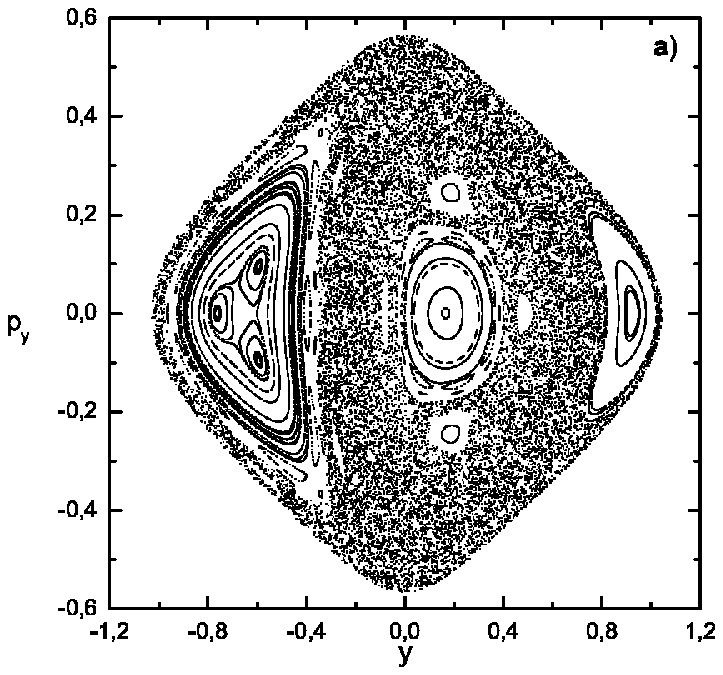}\hspace{-0.85cm}
\includegraphics[height=6cm]{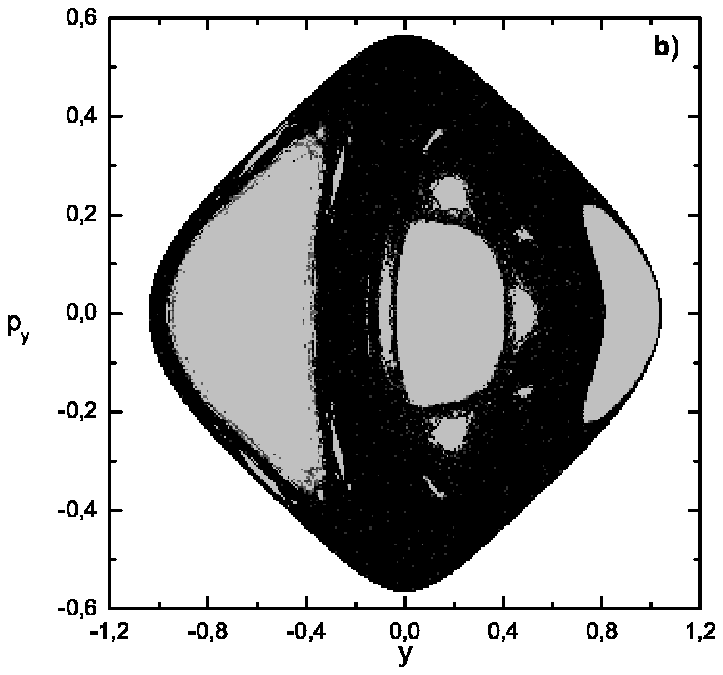}
\caption{Agreement between the results of the Poincar\'e surface of
section (PSS) and SALI. a) The Poincar\'{e} surface of section for
the 2D Ferrers model and $H=-0.335$. b) Regions of different values
of the SALI for 50,000 initial conditions on the $(y,p_{y})$--plane
integrated up to $t=5000$ for the same value of the Hamiltonian. The
light grey colored areas correspond to regular orbits, while the
dark black ones to chaotic. Few gray and dark gray colored points
correspond to ``sticky" orbits. Note the excellent agreement between
the two methods as far as the gross features are concerned, as well
as the fact the SALI can easily pick out small regions of
instability which the PSS has difficulties detecting.}
\label{pss_scan}\vspace{-0.5 cm}
\end{figure}
\vspace{-0.5 cm}
\section{The model}
\label{Model_gal}

A 3D rotating  model of a barred galaxy can be described by the
Hamiltonian function:

\begin{equation}\label{eq:Hamilton}
   H=\frac{1}{2} (p_{x}^{2}+p_{y}^{2}+p_{z}^{2})+ V(x,y,z) -
   \Omega_{b} (xp_{y}-yp_{x}).
\end{equation}

The bar rotates around its $z$ axis, while the $x$ axis is along its
major axis and the $y$ axis is along its intermediate axis. The
$p_{x},p_{y}$ and $p_{z}$ are the canonically conjugate momenta.
Finally, $V$ is the potential, $\Omega_{b}$ represents the pattern
speed of the bar and $H$ is the total energy of the system in the
rotating frame of reference (Jacobi constant). The corresponding
equations of motion are:

\begin{flalign}\label{eq_motion}
 \dot{x}& =  p_{x} + \Omega_{b} y,& \quad  \dot{y}& =  p_{y} - \Omega_{b}x,&    \dot{z}& = p_{z},& \\
 \dot{p_{x}}& =  -\frac{\partial V}{\partial x} + \Omega_{b} p_{y},&  \dot{p_{y}}& = -\frac{\partial V}{\partial y} - \Omega_{b} p_{x},&  \dot{p_{z}}& = -\frac{\partial V}{\partial z}.& \nonumber
\end{flalign}

The equations of the evolution of the deviation vectors and the
calculation of the SALI are given by the corresponding variational
equations.

The potential $V$ of our model consists of three components:

\begin{enumerate}
 \item A \textit{disc}, represented by a Miyamoto potential \cite{Miy.1975}:
\begin{equation}\label{Miy_disc}
  V_D=- \frac{GM_{D}}{\sqrt{x^{2}+y^{2}+(A+\sqrt{z^{2}+B^{2}})^{2}}},
\end{equation}

where $M_{D}$ is the total mass of the disc, $A$ and $B$ are the
horizontal and vertical scale lengths and $G$ is the gravitational
constant.

\item A \textit{bulge}, which is modeled by a Plummer sphere whose
potential is:
\begin{equation}\label{Plum_sphere}
    V_S=-\frac{G M_{S}}{\sqrt{x^{2}+y^{2}+z^{2}+\epsilon_{s}^{2}}},
\end{equation}
where $\epsilon_{s}$ is the scale length of the bulge and $M_{S}$ is
its total mass.

\item A triaxial Ferrers \textit{bar}, the density $\rho(x)$ of which
is:
\begin{equation}\label{Ferr_bar}
  \rho(x)=\begin{cases}\rho_{c}(1-m^{2})^{2} &, m<1  \\
              \qquad 0 &, m\geq1 \end{cases},
\end{equation}
where $\rho_{c}=\frac{105}{32\pi}\frac{G M_{B}}{abc}$ is the central
density, $M_{B}$ is the total mass of the bar and
\begin{equation}\label{Ferr_m}
  m^{2}=\frac{x^{2}}{a^{2}}+\frac{y^{2}}{b^{2}}+\frac{z^{2}}{c^{2}},
\qquad a>b>c> 0,
\end{equation}
with $a,b$ and $c$ being the semi--axes. The corresponding potential
is:
\begin{equation}\label{Ferr_pot}
    V_{B}= -\pi Gabc \frac{\rho_{c}}{n+1}\int_{\lambda}^{\infty}
    \frac{du}{\Delta (u)} (1-m^{2}(u))^{n+1},
\end{equation}
where
\begin{equation}\label{mu2}
m^{2}(u)=\frac{x^{2}}{a^{2}+u}+\frac{y^{2}}{b^{2}+u}+\frac{z^{2}}{c^{2}+u},
\end{equation}

\begin{equation}\label{Delta}
\Delta^{2} (u)=({a^{2}+u})({b^{2}+u})({c^{2}+u}),
\end{equation}
$n$ is a positive integer (with $n=2$ for our model) and $\lambda$
is the unique positive solution of:
\begin{equation}\label{mu2_lamda}
    m^{2}(\lambda)=1,
\end{equation}
outside of bar ($m\geq 1$) and $\lambda=0$ inside the bar.
\end{enumerate}
This model has been used extensively for orbital studies
\cite{Pfe,PSA02,PSA03a,PSA03b,SPA02a,SPA02b} and we will refere to it
hereafter as the Fererrs model. We adopt the following
values of parameters:\ $G=1$, $\Omega_{b}=0.054$ (54 $\frac{Km}{sec
\cdot Kpc}$),  $a=6$, $b=1.5$, $c=0.6$, $A=3$, $B=1$,
$\epsilon_{s}=0.4$, $M_{B}=0.1$, $M_{S}=0.08$, $M_{D}=0.82$. The
units we use, are: 1 kpc (length), 1 Myr (time) and $2 \times
10^{11} M_{\bigodot}$ solar masses (mass). The
total mass $G(M_{S}+M_{D}+M_{B})$ is set to be equal to 1.\\

\begin{figure}[t]
\centering
\includegraphics[height=6cm]{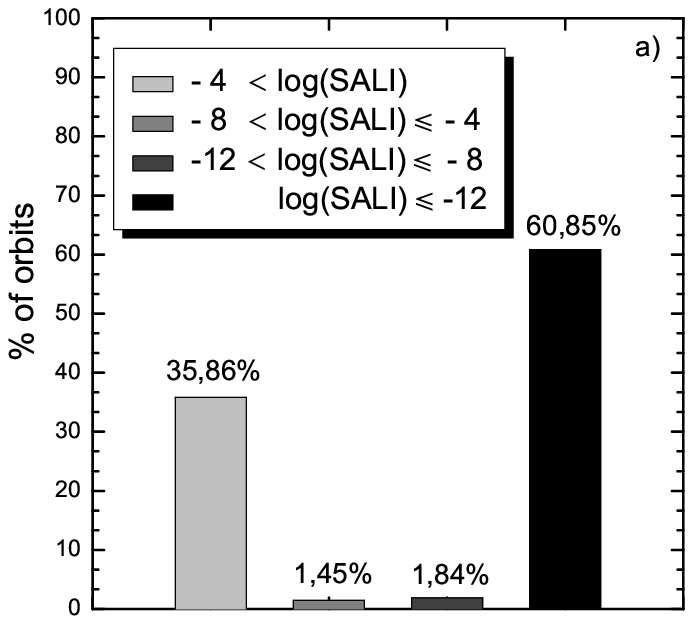}\hspace{-0.8cm}
\includegraphics[height=6cm]{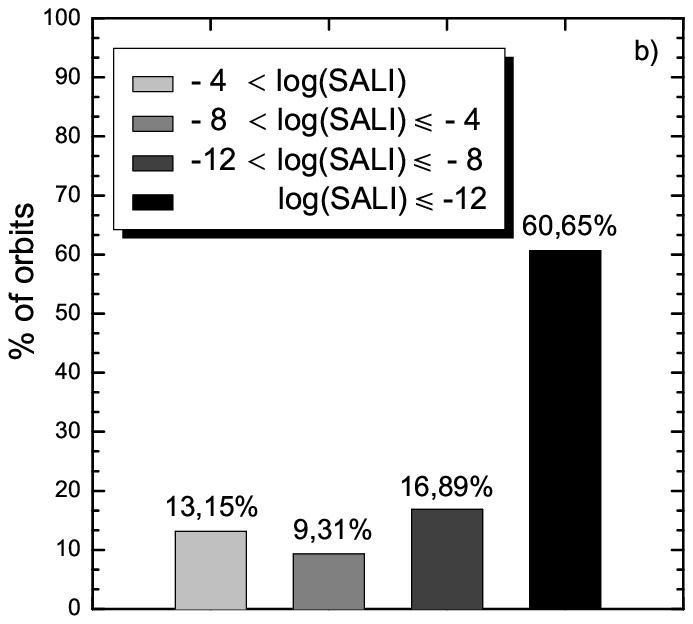}
\caption{a) Percentages of regular (light gray bar), chaotic (black
bar) and sticky orbits (two intermediate bars -- gray and dark gray
bars) for the 2D Ferrers model from a mesh of 50,000 orbits on the
$(y,p_{y})$ plane of Fig. 1 and final time of integration $t=5000$.
b) The percentages for 3D, using 50,000 orbits and the same
classification. The initial conditions are taken as described in the
text.} \label{perc_A}\vspace{-0.5 cm}
\end{figure}

\section{Results in the 2D and 3D Ferrers model} \label{2dof}

The 2D Ferrers model is a subcase of the general 3D one and it can
be described by the Hamiltonian equation (\ref{eq:Hamilton}) by
setting $(z,p_z)=(0,0)$:
\begin{equation}\label{eq:2dof}
   H=\frac{1}{2} (p_{x}^{2}+p_{y}^{2})+ V(x,y) - \Omega_{b} (xp_{y}-yp_{x}).
\end{equation}
Fixing $H=-0.335$, $x=0$, we chose 50,000 initial conditions on the
$(y,p_{y})$--plane of the Poincar\'{e} surface of section (PSS),
while $p_x=H(x,y,p_y)$ and we calculated the final values of the
SALI for $t=5000$. We were able to detect very small regions of
instability that can not be visualized easily by the PSS method. In
Fig.~\ref{pss_scan}, on the left panel, we have plotted the PSS for
this Hamiltonian value. On the right panel we attributed to each
grid point a color according to the value of the SALI at the end of
the evolution. The light grey color corresponds to regular orbits
and to the areas that host them while the black color represents the
chaotic ones. The intermediate colors between the two extremes
represent the so--called ``sticky" orbits, i.e. orbits whose nature
is chaotic but need more time to show their behavior (weak-chaotic
orbits). The distinction between them is done by measuring SALI at
$t=5000$: orbits with SALI$\leq10^{-12}$ correspond to strongly
chaotic, orbits with SALI$>10^{-4}$ to regular, while orbits with
$10^{-8}<$SALI$\leq10^{-4}$ and $10^{-12}<$SALI$\leq10^{-8}$ are
``sticky". In Fig.~\ref{perc_A}a we present the percentages of
orbits from this PSS, according to this classification. We find that $\simeq
37,3\%$ of the orbits on this PSS are regular.

For the 3D case of the model we used a sample of 50,000 orbits,
equally spaced on a cartesian grid in the $(x,z,p_{y})$ space, with $x\in
[0.0,7.0]$, $z\in [0.0,1.5]$ and $p_y \in [0.0,0.45]$, while
$(y,p_{x},p_{z})=(0,0,0)$. In this way, we attempted to create
initial conditions that could support the bar. These orbits
could be parented mainly by the $x_{1}$ tree, i.e. the $x_{1}$ family and
the $x_1v_i$ $i = 1,...$ families that bifurcate from it and extend
vertically well
above the disc region \cite{PSA02}. Although these
initial conditions cover all the available energy interval, they are
not spread uniformly over it. Note also that a few of
these have an energy value beyond the escape energy, and we dismiss
them. These, however, constitute less than $0.5\%$ of the total, so that they
influence very little our study and the statistics.\
\begin{figure}[t]
\centering
\includegraphics[height=6cm]{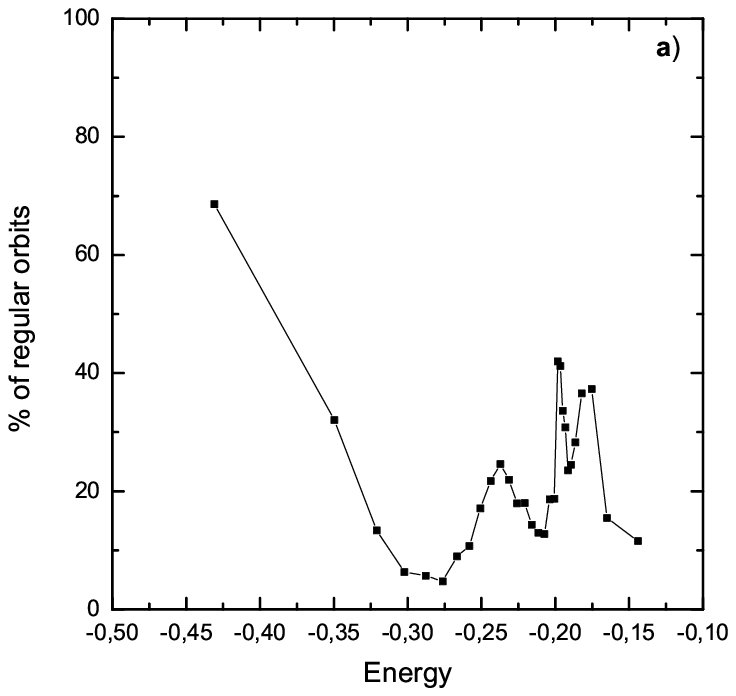}\hspace{-0.8cm}
\includegraphics[height=6cm]{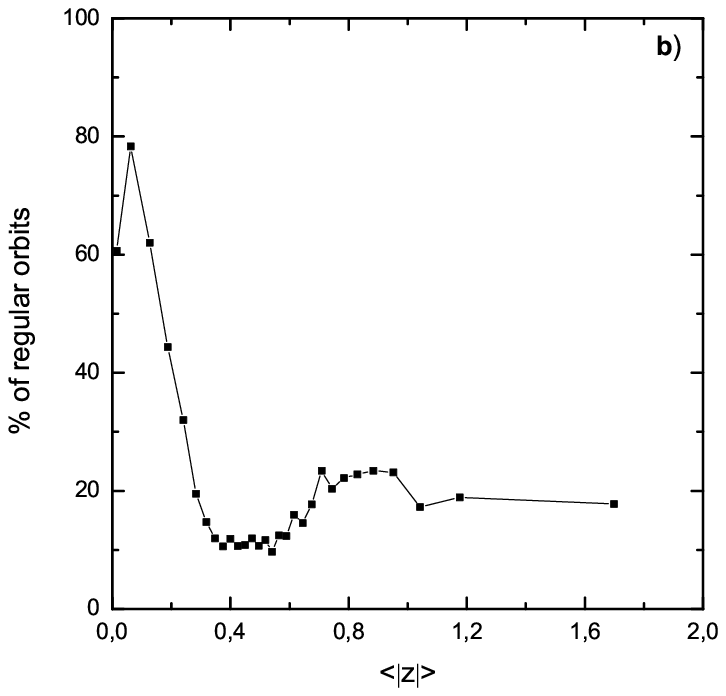}\vspace{-0.5cm}
\includegraphics[height=6cm]{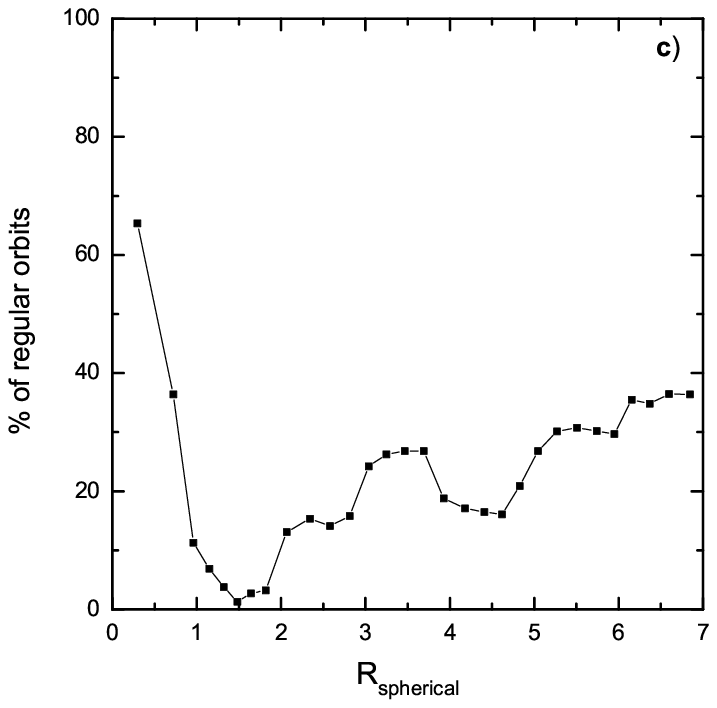}\hspace{-0.8cm}
\includegraphics[height=6cm]{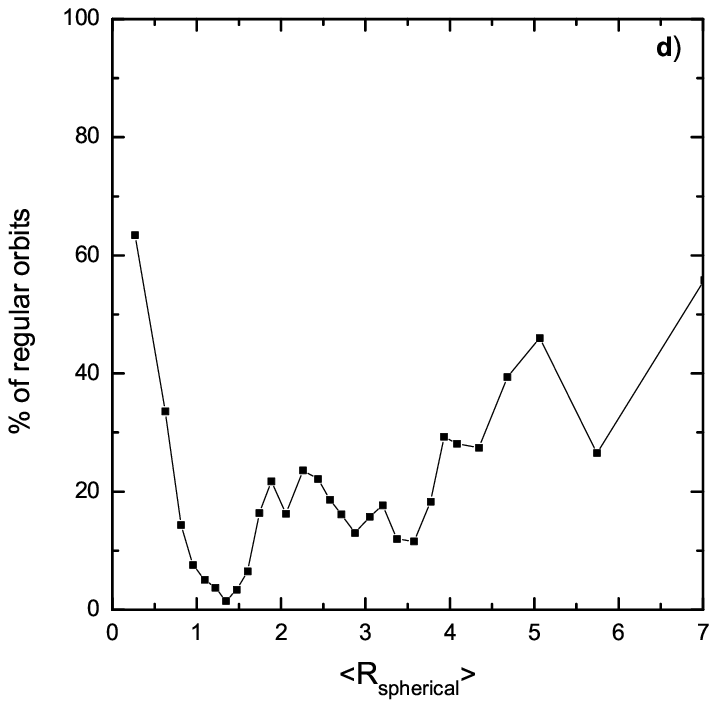}
\caption{Percentages of regular orbits as a function of: the Jacobi
  constant
(panel a), the absolute value of their $z$ coordinate, $<\mid z
\mid>$, meaned over their orbital evolution (panel b), their initial
spherical radius, $R_{spherical}$ (panel c) and the spherical
radius, $<R_{spherical}>$, meaned over their motion (panel d).}
\label{perce}\vspace{-0.5 cm}
\end{figure}
In Fig.~\ref{perc_A}b we present the corresponding percentages of
regular, chaotic and sticky orbits. We find that, using these
initial conditions and parameters, the phase space of the model is
dominated by chaotic orbits, since 77.54$\%$ of them are chaotic.

In order to check the variation of the percentages of regular and
chaotic orbits as the energy of the model varies, we did the
following:\ We first sorted the energy values for all the initial
conditions. Then, we created 30 energy intervals containing equal
number of orbits, since our way of giving initial conditions does
not imply their uniform distribution in the total energy interval.
In every energy interval we calculated the percentages of regular
and chaotic orbits, considering as chaotic all the orbits with
SALI$<10^{-8}$. In Fig.~\ref{perce}a we show the percentage of
regular orbits in each energy interval, as a function of the mean
energy in that interval.
Generally, the percentage of regular orbits decreases as the energy
increases, but before and after the escape energy (where the Jacobi
constant value is $H \simeq -0.20$) there are two peaks. This
non--monotonic behavior is related to the appearance or
disappearance of stable periodic orbits in the phase space and the
size variation of the stability regions around them.

We also attempted to explore the way that regular and chaotic orbits
are distributed along the $z$--direction of the configuration space.
Following the evolution of each orbit, we calculated the mean of the
absolute value of their $z$ coordinate ($<\mid z \mid>$). Then, we
divided the available $<\mid z \mid>$--interval in 30 slices with
equal number of orbits in each one of them. This restriction gives
us better samples for the estimation of the percentages, implying at
the same time that these slices are not equally sized necessarily.
For every slice separately we calculated the fraction of regular
orbits and in Fig.~\ref{perce}b we plot these percentages as a
function of the $<\mid z \mid>$ in that slice. It
reveals that the slices `near' the $(x,y)$--plane ($<\mid z \mid>
\quad < 0.35$) contain mainly `regular' orbits. Contrarily, slices
for larger values of $z$ host mainly chaotic motion.

Furthermore, we looked at these percentages as a function of the
initial spherical radius ($R_{spherical}$) and the mean spherical
radius ($<R_{spherical}>$, meaned over the evolution). Again, dividing in
30 slices the total range of the $R_{spherical}$, in a similar
manner with the $<\mid z \mid>$, we calculated the percentages of
regular orbits. We plot this for every slice, as a function of
the mean $R_{spherical}$ of that slice. We see that the fraction of
regular orbits decreases strongly with increasing $R_{spherical}$ up to
$R_{spherical}<1.5$ where it reaches a minimum, while for
$1.5<R_{spherical}<7$ this percentage starts increasing gradually.
This result is in good agreement with the results in
Fig.~\ref{perce}d, where the horizontal axis corresponds to the
value of the spherical radius meaned over time during the evolution.
\vspace{-0.5 cm}
\section{Conclusions}
\label{Conclusions}

We used SALI to study the dynamical behavior of Hamiltonian models
of 2D and 3D barred galaxies. We found that in both cases there is a
significant amount of chaotic orbits. In the 2D model, we were able
to chart a subspace of the phase space, to identify rapidly even
tiny regions of regular motion and measure their percentages. In the
3D model, apart from computing the global percentages of regular and
chaotic orbits, we calculated these percentages as a function of the
energy and found that low values of the energy are mainly dominated
by `regular' orbital motion. We also followed the distribution of
the chaotic and regular orbits in the configuration space and we
found that orbits which lie near the $(x,y)$--plane with relatively
small mean deviations in $z$--direction are generally regular.
Finally, we monitored the variation of their percentages as a function of
their initial spherical radius and their mean spherical radius. We
find that the fraction of regular orbits is dominant in
regions near the center, as well as at relatively larger distances from it.

\vspace{-0.5 cm}
\section{Acknowledgments}
T.~Manos was partially supported by the ``Karatheodory" graduate
student fellowship No B395 of the University of Patras, the program
``Pythagoras II" and the Marie Curie fellowship No
HPMT-CT-2001-00338. We acknowledge financial support from grant
ANR-06-BLAN-0172.



\end{document}